\PassOptionsToPackage{usenames,dvipsnames,svgnames,table}{xcolor}
\documentclass[9pt]{entcs}
\usepackage{entcsmacro}
\usepackage{graphicx}
\usepackage{tikz,tikz-cd}
\usepackage{okcategory,oksemantics}
\usepackage{multicol}
\def\lastname{Kammar and McDermott}

\volume{Under consideration}

\journal{Under consideration}

\def\firstheadline{\vbox{\baselineskip=10pt \parindent 0pt \obeylines
    \eightrm\hfill Under consideration\hfill\break}}
\def\firstfootline{\vbox{\baselineskip=10pt \parindent 0pt \obeylines
    \vspace{.1in}\centerline{\eightit
      } \centerline{\eightit
      }
    \centerline{}}}

\usepackage{todonotes}
\usepackage{hyperref}

\newcommand\inpar[1]{\par\noindent\textbf{#1.}}
\newcommand\ML{\textsc{ml}}
\newcommand\wcpo{$\omega$-cpo}

\newcommand\wchain{$\omega$-chain}
\newcommand\wchains{\wchain{}s}
\newcommand\return{\mathop{\mathrm{return}}\nolimits}
\newcommand\mult\mu
\newcommand\bind{\mathrel{\scalebox{.5}[1]{$>\!\!>\!=$}}}
\renewcommand\alg{\mathrm{alg}}
\newcommand\uncurry{\mathop{\mathrm{uncurry}}\nolimits}
\newcommand\epi[1]{#1^{\mathbf{e}}}
\newcommand\mono[1]{#1^{\mathbf{m}}}
\usepackage{mathpartir}
\usepackage{listings}
\usepackage{xcolor}
\lstdefinelanguage{calculus}{
  morekeywords={let, in},
  keywordstyle=\bfseries\color{DarkBlue},
  morekeywords=[2]{True, False},
  keywordstyle=[2]\color{Magenta},
  morekeywords=[3]{nat, bool, unit, int, loc, value},
  keywordstyle=[3]\color{DarkGreen},
  morecomment=[l]{--},
  morekeywords=[4]{get, :=},
  keywordstyle=[4]\color{DarkRed},
  commentstyle=\color{gray},
  mathescape=true,
  escapebegin={\color{DarkGreen}},
  literate={%
    {>=}{{$\geq$}}1%
    {<=}{{$\leq$}}1%
    {&&}{{$\land$}}1%
    {\\}{{$\lambda$}}1%
    {->}{{$\to$}}2%
    {-e->}{{$\xto{\epsilon}$}}2%
    {_}{{\_}}1%
    {lop}{{$\ell^\prime$}}1%
    {loc}{{$\ell$}}1%
    {ref}{{\color{DarkGreen}ref}}3%
    {Loc}{{\color{DarkGreen}loc}}3%
    {A}{{$A$}}1%
    {B}{{$B$}}1%
    {f}{{$f$}}1%
  }%
}

\newcommand\getop{\mathrm{get}}
\newcommand\setop{\mathrm{set}}

\lstnewenvironment{calculus}{\lstset{language=calculus, xleftmargin=.2\textwidth}}{}
\lstnewenvironment{calculus*}[1][1]{%
  \lstset{language=calculus, numbers=left, numberstyle=\tiny, firstnumber=#1, stepnumber=1, xleftmargin=.2\textwidth}%
  \center
}{%
  \endcenter
}

\newcommand\calc[1]{\lstinline[language=calculus]{#1}}
\newcommand\C{\Cat C}
\newcommand\D{\Cat D}
\newcommand\E{\Cat E}
\newcommand\M{\Cat M}
\newcommand\W{\Cat W}

\newcommand\Rightarrowdot{\dot{\Rightarrow}}
\renewcommand\exp[2]{{#1}\mathbin{\Rightarrow}{#2}}
\newcommand\expdot[2]{{#1}\mathbin{\Rightarrowdot}{#2}}

\newcommand{\sem}{\scottbrackets}
\newcommand\strength{\mathrm{str}}
\newcommand\Mnd[2]{{#2}\operatorname{-\Category{Mnd}^{#1}}}
\newcommand\op{\mathrm{op}}
\newcommand{\functors}[2]{[{#1},{#2}]}
\newcommand\colim{\mathop{\mathrm{colim}}}
\newcommand\rightarrowdot{\xrightarrow{.}}
\newcommand\Fibre[2]{{#1}_{#2}}
\newcommand\direct[1]{{#1}_*}
\newcommand\inverse[1]{{#1}^*}
\newcommand\cod{\mathrm{cod}}
\newcommand\LogRel{\Category{LogRel}}

\newcommand\lambdac{\ensuremath{\lambda_{\text{c}}}}

\newcommand\syntax[1]{\mathsf{#1}}
\newcommand\unittype{\syntax{1}}
\newcommand\emptytype{\syntax{0}}
\newcommand\ttterm{()}
\newcommand\fstterm[1]{\syntax{fst}\,{#1}}
\newcommand\sndterm[1]{\syntax{snd}\,{#1}}
\newcommand\botelimterm[1]{\syntax{elim}_\emptytype\,{#1}}
\newcommand\inlterm[1]{\syntax{inl}\,{#1}}
\newcommand\inrterm[1]{\syntax{inr}\,{#1}}
\newcommand\caseterm[5]{\syntax{match}~{#1}~\syntax{with}~\{\inlterm{#2}.\,{#3}, \inrterm{#4}.\,{#5}\}}

\newcommand\erase[1]{\underline{#1}}

\newcommand\BaseType{\mathcal{B}}
\newcommand\Constant{\mathcal{K}}
\newcommand\welltyped[3]{{#1} \vdash {#2} : {#3}}
\newcommand\effwelltyped[4]{{#1} \vdash_{#4} {#2} : {#3}}

\newcommand\coend[2]{\int^{#1} #2}
\newcommand\lookup{\mathrm{lookup}}
\newcommand\update{\mathrm{update}}
\newcommand\alloc{\mathrm{alloc}}
\newcommand\reftype{\calc{ref}}
\newcommand\valuetype{\calc{value}}
\newcommand\I{\mathbb{I}}
\newcommand\Loc{\mathbb{L}}
\newcommand\State{\mathbb{S}}
\usepackage{bbold}
\renewcommand\terminal{\mathbb{1}}
\newcommand\V{\mathbb{V}}

\begin{document}
\begin{frontmatter}
  \title{Factorisation systems for logical relations and monadic
    lifting in type-and-effect system semantics}
  \author{Ohad Kammar\thanksref{ohad}}
  \address{%
    Department of Computer Science   \\
    University of Oxford             \\
    Oxford, England}
  \author{Dylan McDermott\thanksref{dylan}}
  \address{%
    Computer Laboratory \\
    University of Cambridge \\
    Cambridge, England%
  }
  \thanks[ohad]{Email:
    \href{mailto:ohad.kammar@cs.ox.ac.uk}{\texttt{\normalshape
        ohad.kammar@cs.ox.ac.uk}}%
  }
  \thanks[dylan]{Email:
    \href{mailto:dylan.mcdermott@cl.cam.ac.uk}{\texttt{\normalshape
        dylan.mcdermott@cl.cam.ac.uk}}%
  }
  \begin{abstract}
    Type-and-effect systems incorporate information about the
    computational effects, e.g., state mutation, probabilistic choice,
    or I/O, a program phrase may invoke alongside its return value. A
    semantics for type-and-effect systems involves a parameterised
    family of monads whose size is exponential in the number of
    effects. We derive such refined semantics from a single monad over
    a category, a choice of algebraic operations for this monad, and a
    suitable factorisation system over this category. We relate the
    derived semantics to the original semantics using fibrations for
    logical relations. Our proof uses a folklore technique for lifting
    monads with operations.
  \end{abstract}
\end{frontmatter}

\section{Introduction}
Consider the following program phrase in an imperative-functional \ML-like language:
\begin{calculus*}
let (triple:unit->int) = \_:unit. 3*(get loc)
in loc := 1;
   loc := triple() + triple()
\end{calculus*}
The locally-defined function \calc{triple:unit->int} triples the
value read from memory location \calc{loc}. The phrase then triples
this value twice, and mutates the state to the sum of these two
results. When optimising the program, we would like to cache the call
to \calc{triple}, and replace line~3 with a single memory access:
\begin{calculus*}[3]
   loc := let y = triple() in y + y
\end{calculus*}
This transformation only preserves the semantics because the
computational effects \calc{triple} invokes are limited to reading. If
we replace instead its definition on line~1 with a function that
increments location \calc{lop} with each invocation of \calc{triple}
then the caching optimisation is no longer semantics preserving:
\begin{calculus*}[1]
let (triple:unit->int) = \_:unit. lop := (1 + get lop); 3*(get loc)
\end{calculus*}
\emph{Type-and-effect
  systems}~\cite{lucassen-gifford:polymorphic-effect-systems} refine
types, such as \calc{triple:unit->int}, to propagate the information
about which computational effects code pieces may invoke, e.g.,
decorating function types with additional \emph{effect annotations}:
\begin{calculus}
triple:unit-e->int
\end{calculus}
In \emph{Gifford-style} systems, these annotations are finite sets of
effect operations, such as $\epsilon := \set{\getop, \setop}$. For
example, for every proper subset $\epsilon \subsetneq \set{\getop,
  \setop}$, the caching transformation for every function $\calc{f : A
  -e-> B}$ is semantics preserving, while for $\epsilon = \set{\getop,
  \setop}$ it is not.

Adequate denotational semantics is a natural technique for validating
such equational transformations, and there is a long line of work
validating type-and-effect-dependent transformations, starting with
independent results by Tolmach~\cite{tolmach:hierarchy},
Wadler~\cite{Wadler:1998:MEM:291251.289429}, and Benton et
al.~\cite{benton-kennedy-russell:compiling-standard-ml-to-java-bytecodes},
and continuing to this day~\cite{BENTON201827}. In their most general
form, the semantics for an effect system consists of a \emph{graded
  monad}~\cite{Katsumata:2014:PEM:2578855.2535846}, a compatible
family of monad-like structures $T_\epsilon$ indexed by the effect
annotations $\epsilon$.

Here we make two contributions:

\inpar{Contribution 1: avoiding structural combinatorial blow-up}
To give the model structure for an arbitrary Gifford-style
type-and-effect system with $n$ operation symbols, one would need to
give the structure of $2^n$ different monad-like structures,
$n2^{n-1}$ monad-like-morphisms, and commute more than the same amount
of diagrams to discharge the relevant proof obligations. To circumvent
this blow-up, for example, Benton et al. give uniform bespoke
definitions for each $T_\epsilon$, e.g., as in \cite{BENTON201827}. To
avoid an ad-hoc definition for each collection of effects,
Katsumata~\cite{Katsumata:2014:PEM:2578855.2535846} constructs graded
monads for Gifford-style systems when the effects in the language are
free. Here we give a general construction for Gifford-style systems
whose effects are given by a set of Kleisli arrows for an arbitrary
monad over a category with a factorisation system with appropriate
closure properties, providing a uniform construction even when the
effects of interest are not free.

\inpar{Contribution 2: relationship to a base semantics}
We also show that this construction gives sound and complete reasoning
principles with respect to the original semantics under additional
assumptions. As usual, such proofs involve constructing a logical
relation. Here, we work fibrationally using Katsumata's notion of a
\emph{fibration for logical
  relations}~\cite{katsumata:relating-computational-effects}. We
extend Hughes and Jacobs's characterisation of fibrations of
factorisation
systems~\cite{hughes-jacobs:factorization-systems-fibrations} and
characterise the factorisation systems that correspond to fibrations
for logical relations. Finally, we also define generally a monadic
lifting for an arbitrary monad along a fibration for logical relations
that also lifts a given collection of Kleisli arrows. This
construction utilises the bijection between algebraic operations and
generic
effects~\cite{plotkin-power:algebraic-operations-generic-effects}.
While Kammar~\cite{DBLP:phd/ethos/Kammar14} describes it in the
special set-theoretic case, we believe this folklore monadic lifting
methodology\footnotemark{} should be known in its greater
generality. We demonstrate that our results are applicable in several
cases of interest.
\footnotetext{Alex K.~Simpson, private communication, 2015.}

These two contributions substantially generalise Kammar and Plotkin's
previous domain-theoretic~\cite{Kammar:2012:AFE:2103656.2103698} and
set-theoretic constructions~\cite{DBLP:phd/ethos/Kammar14}. Our
factorisation system construction also strictly generalises the one in
Kammar's thesis~\cite{DBLP:phd/ethos/Kammar14}, which is limited to
factorisation of enriched Lawvere theories~\cite{power1999enriched}
over a locally presentable category. The development here is also
substantially simpler than Kammar's thesis. This simplification occurs
in two levels. Kammar's previous development requires a combinatorial
solution set condition argument using Bousfield's factorisation
theorem~\cite{bousfield:constructions-of-factorization-systems-in-categories},
while our factorisation construction is structural and
elementary. Second, our proofs are straightforward in comparison.

The rest of the paper is structured as
follows. Section~\ref{sec:factorisation} presents our main
factorisation construction. Section~\ref{sec:calculus} uses this
construction to give semantics for a type-and-effect system for
Moggi's computational $\lambda$-calculus. Section~\ref{sec:lifting}
instruments a logical relations soundness and completeness proof from
the factorisation construction. Section~\ref{sec:examples} surveys
example applications to our construction. Section~\ref{sec:conclusion}
concludes.

\section{Factorising monads}\label{sec:factorisation}
To present our main construction, we first review the relevant
category theoretic concepts and results.

\subsection{Preliminaries and terminology}

We assume familiarity with category theory, including categories $\C$,
$\D$, functors $F, G : \C \to \D$, and natural transformations
$\alpha, \beta : F \to G$, and related concepts as found in textbooks
such as Mac Lane's~\cite{maclane:working-mathematician}.

\subsubsection{Factorization systems}

A factorisation system axiomatises the set-theoretic situation in
which every function $f : A \to B$ can be factorised as $f = m \compose
e$, i.e., a surjection $e : A \epimor f[A]$ onto the image of $f$,
followed by the injection $m : f[A] \monomo B$ of this image into
$f$'s codomain. In the general situation, we have two classes of
morphisms $\pair\E\M$ over a category $\C$, where $\E$-morphisms are
thought of as epimorphisms and $\M$-morphisms are thought of as
monomorphisms. We adopt the common convention to reserve the notation
$e : A \epimor B$ for an $\E$-morphism and $m : B \monomo C$ for an
$\M$-morphism when $\E$ and $\M$ are clear from the context, but
emphasise that neither class needs to consist of epis or monos.

\label{section:factorization-systems}
\begin{definition}
  An \emph{orthogonal factorisation system} on a category $\C$ is a pair
  $\pair{\E}{\M}$ consisting of two classes of morphisms of $\C$ such
  that:
  \begin{itemize}
  \item
    Both $\E$ and $\M$ are closed under composition, and contain
    all isomorphisms.
  \item
    Every morphism $f : X \to Y$ in $\C$ factors into $f = m \compose e$
    for some $m \in \M$ and $e \in \E$.
  \item
    The \emph{diagonal fill-in} property is satisfied: for each
    commutative square as on the left, with $m \in \M$ and $e \in
    \E$ there is a unique morphism $h : X \to Y$ such that
    $h\compose m = f$ and $e\compose h = g$, as on the left below:
    \begin{center}
      \begin{tikzcd}
        W \arrow[r, "e", two heads]
        \arrow[d, "f" left] &
        X \arrow[d, "g"]
        \arrow[ld, "=", phantom]
        \\
        Y \arrow[r, "m" below, tail] &
        Z
      \end{tikzcd}
      $\implies$
      \begin{tikzcd}
        |[alias=W]| W \arrow[r, "e", two heads]
        \arrow[d, "f" left] &
        X \arrow[d, "g"]
        \arrow[ld, "h"{name=h} below, dashed]
        \\
        Y \arrow[r, "m" below, tail] &
        Z \arrow["=", phantom, to=h]
        \arrow["=", phantom, from=W, to=h]
      \end{tikzcd}
      \hfil
      \begin{tikzcd}[row sep=scriptsize]
        X \arrow[r, "f" above]
        \arrow[d, "g_1" left] &
        Y
        \arrow[d, "g_2" right]
        \\
        X' \arrow[r, "f'" below] &
        Y'
        \arrow[ul, "=", phantom]
      \end{tikzcd}
      $\implies$
      \begin{tikzcd}[row sep=scriptsize]
        X \arrow[r, "e" above, two heads]
        \arrow[d, "g_1" left]
        \arrow[rd, "=", phantom] &
        A \arrow[r, "m" above, tail]
        \arrow[d, "{h}" , dashed]
        &
        Y
        \arrow[d, "g_2" right]
        \\
        X' \arrow[r, "e'" below, two heads] &
        A' \arrow[r, "m'" below, tail]
        &
        Y'
        \arrow[ul, "=", phantom] &
      \end{tikzcd}
    \end{center}
  \end{itemize}
\end{definition}
Under the first two axioms, the diagonal fill-in axiom is equivalent
to a form of functoriality in factorisation, as in the implication
above on the right.  In addition, it implies that factorisations of
morphisms are unique up to a unique canonical iso, and so we talk
about \emph{the} factorisation of a morphism.

\begin{example}
  The category $\Set$ has (surjection, injection) as a factorisation
  system, i.e., $\E$ is the class of surjective functions and $\M$ is
  the class of injective functions.
\end{example}
\begin{example}[Meseguer~\cite{meseguer:factorisations}]
  Consider the category $\wCPO$ of partial orders possessing all least
  upper bounds (lubs) of $\omega$-indexed monotone sequences
  (\wchains), i.e., \wcpo{}s, and monotone functions between them
  preserving these lubs, i.e., Scott-continuous functions. A
  \emph{dense function} is a continuous function $e : X
  \twoheadrightarrow Y$ such that the smallest \wchain-closed
  subset $U\subseteq Y$ with $e[X] \subseteq U$ is $Y$ itself, i.e., a
  Scott-continuous function with a dense image. A \emph{full function}
  is a continuous function $m : X \rightarrowtail Y$ such that $m\,x
  \leq m\,x'$ implies $x \leq x'$ for each $x \in X$. The category
  $\wCPO$ has (dense, full) as a factorisation system. Every full
  function is necessarily injective, but there are non-surjective
  dense functions~\cite{lehmann-pasztor:epis-need-not-be-dense}.
\end{example}
\begin{example}
  Consider the functor category $\functors{\W}{\C}$, for a small
  category $\W$ and any category $\C$, and let $\pair{\E}{\M}$ be a
  factorisation system on $\C$. Take $\E'$ (respectively $\M'$) as the
  class of natural transformations that are component-wise in $\E$
  (respectively $\M$). Then $\pair{\E'}{\M'}$ is a factorisation
  system for $\functors{\W}{\C}$.
\end{example}

The left and right classes in a factorisation system have useful
closure properties. For example, if $g \compose f$ and $f$ are in
$\E$, then so is $g$. For another example, view both classes as full
subcategories of the arrow category $\C^{\rightarrow}$ whose objects
are triples $f = \triple* {A^f_1}{A^f_2}{\underline f}$ consisting of
a morphism $\underline f : A^f_1 \to A^f_2$, and whose morphisms $h :
f \to g$ are pairs $\pair{h_1}{h_2}$ consisting of morphisms $h_i :
A^f_i \to A^g_i$ making the evident square commute. Then the left
class is closed under colimits in the arrow category, and similarly
the right class is closed under limits.

\subsubsection{Monad structures and monads}\label{subsub:monads}
The main feature of our factorisation construction is its modularity.
First, factorisation takes place on a purely structural level, and we
do need no semantic properties such as the the monad laws. Second,
factorisation takes place on a pay-as-you-go basis, factorising any
additional data the morphism of interest preserves. To describe it
explicitly, we first describe precisely the structures we will
factorise.

A \emph{monad structure} $T$ on a category $\C$ consists of a triple
$\triple{\underline T}{\return^T}{\mult^T}$ where:
\begin{itemize}
\item the \emph{functor part} $\underline T$ assigns to every
  $\C$-object $A$ another $\C$-object $\underline T A$, and to every
  $\C$-morphism $f : A \to B$ another $\C$-morphism $\underline T f :
  \underline TA \to \underline TB$;
\item the \emph{unit} $\return^T$ assigns to every $\C$-object $A$ a
  $\C$-morphism $\return^T_A : A \to \underline TA$; and
\item the \emph{multiplication} $\mult^T$ assigns to every $\C$-object
  $A$ a $\C$-morphism $\mult^T_A : \underline T^2 A \to \underline
  TA$.
\end{itemize}
A \emph{monad} is thus a monad structure $T$ satisfying the well-known
monad laws. When $\C$ has finite products, a \emph{strong monad
  structure} is a monad structure $T$ with an additional structure
component:
\begin{itemize}
\item the \emph{strength} $\strength^T$ assigns to every pair of
  $\C$-objects $A$ and $B$ a $\C$-morphism $\strength^T_{A, B} : A
  \times \underline TB \to \underline T(A\times B)$.
\end{itemize}
A \emph{strong monad} is thus a strong monad structure satisfying the
well-known laws. We similarly define \emph{Kleisli triple structures}
$T = \triple{\underline T}{\return^T}{\bind^T}$, demanding only an
assignment $\underline T$ on object and a Kleisli extension operation
$\bind^T_{A,B}f : \underline TA \to \underline TB$ for every $f : A
\to \underline TB$. Finally, when $\C$ is cartesian closed, we define
a \emph{strong Kleisli triple structure} $T = \triple{\underline
  T}{\return^T}{\bind^T}$ analogously, replacing $\bind$ with an
assignment of a morphism $\bind^T_{A,B}:\underline TA \times
\underline TB^A \to \underline TB$ for every pair of $\C$-objects $A$
and $B$.

Morphisms of such structures $m : S \to T$ assign to every $\C$-object
$A$ a morphism $m_A : \underline SA \to \underline TA$ that preserve
the structure, i.e., satisfy the same conditions a (strong) monad
morphism should. Such morphisms provide super-categories for the
categories of (strong) monads and (strong) Kleisli triples. The usual
isomorphisms between the familiar sub-categories fail to extend to
isomorphisms between the structural super-categories without the
presence of the monad laws.

An \emph{algebra structure} $A = \pair{\underline A}{\alg_A}$ for a
monad structure $T$ over $\C$ consists of:
\begin{itemize}
\item the \emph{carrier} $\underline A$, a $\C$-object; and
\item the \emph{algebra map} $\alg_A$, a $\C$-morphism $\alg_A :
  \underline T\underline A \to \underline A$.
\end{itemize}
When $T$ is a monad, an \emph{algebra} is an algebra structure
satisfying the well-known algebra properties. Similarly, when $T$ is a
Kleisli triple structure, an \emph{algebra structure} $A =
\pair{\underline A}{\bind_A}$ replaces the algebra map with:
\begin{itemize}
\item the \emph{extension operator} $\bind_A$ which assigns to every
  morphism $f : X \to \underline A$ a morphism $\bind f : \underline TX \to
  \underline A$.
\end{itemize}
When $T$ is a Kleisli triple, an \emph{algebra} is an algebra
structure
satisfying~\cite{marmolejo-wood:monads-as-extension-systems}, for
every $f : X \to \underline A$, and every $g : X \to \underline TY$,
$h : Y \to \underline A$:
\begin{center}
\begin{tikzcd}
  X \arrow[r, "\return" above]
  \arrow[dr, "f" below left]
  \arrow[dr, ""{name=f, above right}, phantom]
                               & |[alias=TX]|
                                 \underline TX
                                 \arrow[d, "\bind f" right]
  \\
                               & \underline A
  \arrow["=", phantom, from=f, to=TX]
\end{tikzcd}
\hfil
\begin{tikzcd}
  \underline TX
  \arrow[r, "\bind g" above]
  \arrow[dr, "\bind^A ((\bind^A h) \compose g)" below left]
  \arrow[dr, ""{name=bind, above right}, phantom]
  & |[alias=TY]| \underline TY
    \arrow[d, "\bind h" right]
  \\
  & \underline A
  \arrow["=", phantom, from=TY, to=bind]
\end{tikzcd}
\end{center}
Similarly, we define an algebra structure for a strong Kleisli triple
structure by replacing the extension operator with an internal
extension operator $\bind : \underline TX\times \underline A^X \to
\underline A$, and algebras for a strong Kleisli triple internalise
the two equations above.

Let $T$ be a monad over a category $\C$. Recall that a \emph{Kleisli
  arrow} is a morphism $f : A \to \underline TB$. When $\C$ is
cartesian closed and $T$ is strong, an \emph{algebraic
  operation}~\cite{plotkin-power:algebraic-operations-generic-effects}
$\alpha : A \to B$ for $T$ assigns to every $\C$-object $X$ a
$\C$-morphism $\alpha_X : (\underline TX)^B \to (\underline TX)^A$,
natural in $X$, and respecting the multiplication/extension and the
strength. Plotkin and
Power~\cite{plotkin-power:algebraic-operations-generic-effects}
establish a bijection between Kleisli arrows $f : A \to \underline TB$
and algebraic operations $\alpha : A \to B$ given by:
\[
\uncurry \alpha_X : A\times (\underline TX)^B \xto{f\times\id}
\underline TB \times (\underline TX)^B \xto{\bind} \underline TX
\]

Let $F : \C \rightarrow \C$ be a functor with a tensorial strength
$\strength^F$ over a category with finite products. The category
$\Mnd{\C}{F}$ of $F$-monads on $\C$ has as objects $\pair T\beta$
where $T$ is a strong monad and $\beta : F\compose T \rightarrow T$ is a
natural transformation making the square on the left commute:
\begin{center}
\begin{tikzcd}
        X \times F\,(T\,Y)
        \arrow[r, "\strength^F"]
        \arrow[d, "\id\times \beta" left]
        \arrow[rrd, "=", phantom] &
        F\,(X \times T\,Y)
        \arrow[r, "F\,\strength^T"] &
        F\,(T\,(X\times Y))
        \arrow[d, "\beta"] \\
        X \times T\,Y
        \arrow[rr, "\strength^T" below] & &
        T\,(X\times Y)
\end{tikzcd}
\hfil
\begin{tikzcd}
        F\,(T\,X)
        \arrow[r, "F\,m"]
        \arrow[d, "\beta" left]
        \arrow[rd, "=", phantom] &
        F\,(T'\,X)
        \arrow[d, "\beta'"] \\
        T\,X
        \arrow[r, "m" below] &
        T'\,X
      \end{tikzcd}
\end{center}
A morphisms $m : \pair T\beta \rightarrow \pair{T'}{\beta'}$ consists
of a strong monad morphism $m : T \to T'$ making the square on above
right commute.

An \emph{effect signature} $\epsilon$ in a category $\C$ consists of a
set $\underline \epsilon$ and an $\underline\epsilon$-indexed family
of pairs of $\C$-objects. We write $(\op : X \to Y) \in \epsilon$ when
$\op \in \underline \epsilon$ and $\pair XY$ is the $\op$-th component
in $\epsilon$. We write $\epsilon \subset \epsilon'$ when $\underline
\epsilon \subset \underline\epsilon'$ and both agree
component-wise.

For every effect signature $\epsilon$ we define the functor $F_\epsilon
: \C \to \C$ by
\( F_\epsilon :=
\sum_{(\op : X \rightarrow Y) \in \epsilon} X \times \parent{-}^Y
\).
Every $\Mnd{\C}{F_\epsilon}$-object $\pair T\beta$ induces an
algebraic operation $\alpha_\op$ for each operation in $(\op : X \to
Y) \in \epsilon$, which in turn induces a Kleisli arrow $\pair
T\beta\sem\op : X \to \underline TY$. This process extends to an
isomorphism between $\Mnd{\C}{F_\epsilon}$ and the category whose
objects are \emph{$\epsilon$-monads} on $\C$, i.e., pairs
$\pair{T}{\sem-}$ consisting of a strong monad $T$ together with a
morphism $\sem{\op} : X \rightarrow T\,Y$ for each $(\op : X
\rightarrow Y) \in \epsilon$. Its morphisms $m : \pair T{\sem-}
\rightarrow \pair{T'}{\sem-}$ are strong monad morphisms $m : T \to
T'$ such that, for all $(\op : X \rightarrow Y) \in \epsilon$, we have
\( m \compose \sem\op = \sem\op \).

We recall Kelly's~\cite{%
  kelly:transfinite-constructions,%
  kelly:transfinite-constructions-addenda}
transfinite construction of the free $F$-monad when $\C$ has
$\kappa$-directed colimits and $F$ is \emph{$\kappa$-ranked}, i.e.,
preserves these colimits, for some regular cardinal $\kappa$.  Define
an ordinal-indexed sequence of functors $S_\alpha : \C \to \C$ by
transfinite induction on $\alpha$ as follows:
\begin{align*}
  S_0 &\definedby {\Id} &
  S_{\alpha + 1} &\definedby {\Id} + F\compose S_\alpha &
  S_\lambda &\definedby \colim_{\alpha<\lambda}S_\alpha
  \qquad(\lambda~\text{a limit ordinal})
\end{align*}
Each colimit is directed: the diagram includes morphisms $S_\alpha \to
S_\alpha'$ for $\alpha \leq \alpha'$; each morphism is defined by
transfinite recursion. The free
monad for $F$ is then given by
\( S_F \definedby \colim_{\alpha<\kappa}S_\alpha \).
If $F$ is also strong then $S_F$ is the initial object of
$\Mnd{\C}{F}$.

\subsection{The factorisation theorems}
Let $\pair{\E}{\M}$ be a factorisation system for a category $\C$, and
let $S$ be monad structures on $\C$. We say that $\pair\E\M$ is
\emph{closed under $S$} when, for every $e : A \epimor B$ in $\E$, we
have $\underline S e : \underline S A \epimor \underline SB$ in $\E$. We also
say that $S$ is \emph{compatible} with $\pair\E\M$. In that case, we
can factorise every monad structure morphism $m : S \to T$ through a
monad structure $m[S]$ as a composition of monad structure morphisms
$m : S \epimor m[S] \monomo T$ by choosing a factorisation for each
$m_X$, setting for each $f : X \to Y$, and $Z$:
\begin{center}
\begin{tikzcd}[column sep=3pt, row sep=60pt]
    \underline SX
    \arrow[rr, "m_X"{name=m}]
    \arrow[rd, "\epi m_X" below left, two heads] & &
    \underline TX \\
    &
    \underline{m[S]}X
    \arrow[ru, "\mono m_x" below right, tail]
    \arrow[":=", phantom, from=m]
\end{tikzcd}
\hfil
\begin{tikzcd}[column sep=28pt,row sep=20pt]
  \underline SX
  \arrow[r, "\epi m_X", two heads]
  \arrow[r, ""{name=mx, below}, phantom]
  \arrow[d, "\underline Sf" left]
  & \underline {m[S]}X
  \arrow[d, "\mono m_X" right, tail]
  \arrow[ddl, "\underline{m[S]}f"{name=diagonal, above left}, dashed]
  \arrow[ddl, "" name=diagonal, phantom]
  \\
  \underline SY
  \arrow[d, "\epi m_Y" left, two heads]
  &
  \underline TX
  \arrow[d, "\underline Tf" right]
  \\
  \underline {m[S]}Y
  \arrow[r, "\mono m_Y" below, tail]
  & |[alias=TY]|
  \underline TY
  \arrow["=" above, from=mx, to=diagonal, phantom]
  \arrow["="      , from=TY, to=diagonal, phantom]
\end{tikzcd}
\hfil
\begin{tikzcd}[column sep=3pt, row sep=60pt]
  \underline Z
  \arrow[rr, "\return^{m[S]}_Z"{name=ret, above}]
  \arrow[dr, "\return^S_Z" below left]
  &&
  \underline {m[S]}Z
  \\
  & \underline SZ
  \arrow[ur, "\epi m_Z" below right, two heads]
  \arrow[":=", to=ret, phantom]
\end{tikzcd}
\hfil
\begin{tikzcd}[column sep=18pt, row sep=7pt]
  |[alias=SSZ]|
  \underline S^2Z
  \arrow[r, "\underline S\epi m_Z" above, two heads]
  \arrow[d, "\mult^S_Z" left]
  & \underline S\underline {m[S]}Z
  \arrow[r, "\epi m_{\underline{m[S]}Z}" above, two heads]
  & \underline {m[S]}^2Z
  \arrow[d, "\mono m_Z" right, tail]
  \arrow[dddll,"\mult^{m[S]}_Z" above left, dashed]
  \arrow[dddll,"" name=diagonal, phantom]
  \\
  \underline SZ
  \arrow[dd, "\epi m_Z" left, two heads]
  &&
  \underline{m[S]}\underline TZ
  \arrow[d, "\underline {m[S]}\mono m_Z" right]
  \\
  &&
  \underline T^2Z
  \arrow[d, "\mult^T_Z" right]
  \\
  \underline {m[S]}Z
  \arrow[rr, "\mono m_Z" below, tail]
  &&
  |[alias=TZ]|
  \underline TZ
  \arrow["=", phantom, from=diagonal, to=SSZ]
  \arrow["=", phantom, from=diagonal, to=TZ]
  \end{tikzcd}
\end{center}
This definition makes $\epi m : S \epimor m[S]$ a monad structure
morphism with components in $\E$, and $\mono m : m[S] \monomo T$ a
monad structure morphism with components in $\M$. Using the factorisation
system closure properties, $\pair\E\M$ is also closed under
$m[S]$. Moreover, we have a (component-wise $\E$, component-wise $\M$)
factorisation system of the category of $\pair\E\M$-compatible monad
structures and monad structure morphisms. Every algebra structure $A$
for $m[S]$ induces an algebra structure $S$ by setting:
\[
\alg^T_A : \underline S\underline A \xepimor{\epi m_{\underline A}} \underline{m[S]}\underline A \xto{\alg_A} \underline A
\]

When $\C$ has finite products, we say that a factorisation system
$\pair\E\M$ is \emph{closed under products} when, for every $e_1, e_2
\in E$, we also have that $e_1 \times e_2 \in \E$. We can then
factorise a strong monad structure morphism $m : S \to T$ by setting
the strength for $m[S]$ as on the left:
\begin{center}
  \begin{tabular}[c]{@{}c@{}}
    \begin{tikzcd}[row sep = 18pt, column sep = 40pt]
    |[alias=topleft]|
    X\times \underline SY
    \arrow[r, "\id \times \epi m_Y" above, two heads]
    \arrow[d, "\strength^S_{X,Y}" left]
    &
    X\times \underline{m[S]Y}
    \arrow[d, "\id \times \mono m_Y" right, tail]
    \arrow[ddl, "\strength^{m[S]}_{X, Y}" above left, dashed]
    \arrow[ddl, "" name=diagonal, phantom]
    \\
    \underline S(X \times Y)
    \arrow[d, "\epi m_{X\times Y}" left, two heads]
    &
    X\times \underline TY
    \arrow[d, "\strength^T_{X, Y}" right]
    \\
    \underline{m[S]}(X \times Y)
    \arrow[r, "\mono m_{X\times Y}" below, tail]
    &
    |[alias=botright]|
    T(X \times Y)
    \arrow["=" above left, phantom, from=diagonal, to=topleft]
    \arrow["=", phantom, from=diagonal, to=botright]
    \end{tikzcd}
  \end{tabular}
  \hfil
  \begin{tabular}[c]{@{}c@{}}
  \begin{tikzcd}[row sep = 15pt, column sep=40pt]
    |[alias=upperlt]|
    \underline SX \times \parent{\underline{m[S]}Y}^X
    \arrow[r, "\epi m_X\times\id" above, two heads]
    \arrow[dd, "\bind^S" left]
    &
    \underline{m[S]}X \times \parent{\underline{m[S]}Y}^X
    \arrow[d, "\mono m_X\times \parent{\mono m_Y}^X" right]
    \arrow[ddl, "\bind^{m[S]}" above left, dashed]
    \arrow[ddl, "" name=diagonal, phantom]
    \\
    &
    \underline{T}X \times \parent{\underline TY}^X
    \arrow[d, "\bind^T" right]
    \\
    \underline{m[S]}Y
    \arrow[r, "\mono m_Y" below, tail]
    &
    |[alias=lowerrt]|
    \underline{T}Y
    \arrow["=" above left, phantom, from=diagonal, to=upperlt]
    \arrow["="           , phantom, from=diagonal, to=lowerrt]
  \end{tikzcd}
  \\[-11pt]~
  \end{tabular}
\end{center}
We also include the factorisation construction for strong Kleisli
triples in a cartesian closed category, above on the right. This
construction uses the fact that algebra structures for $m[S]$ induce
algebra structures for $S$.

\begin{theorem}[Factorisation]\label{theorem:monad-factorisation}
  Let $\C$ be a category, $\pair\E\M$ a factorisation system, $S$ and
  $T$ be monads over $\C$, and $m : S \rightarrow T$ a monad
  morphism.
  \begin{itemize}
  \item
    If $\pair\E\M$ is closed under $S$ then $m[S]$ is a monad, and so $\epi
    m$ and $\mono m$ are monad morphisms. As a consequence, every
    algebra for $m[S]$ induces an algebra for $S$.
  \item If, moreover, $\pair\E\M$ is closed under products, $S$, $T$
    are strong monads, and so $m$ is a strong monad morphism, then
    $m[S]$ is a strong monad and $\epi m$, $\mono m$ are strong monad
    morphisms.
  \item When, moreover, $\C$ is cartesian closed, the constructions for
    Kleisli triples and strong monads coincide.
  \end{itemize}
\end{theorem}

The proof, commuting several diagrams, uses the diagonal fill-in
property by substituting definitions.

We can transfer additional structure from $S$ to
$m[S]$. Post-composing with $\epi m$ transfers to $m[S]$ any Kleisli
arrow for $S$. Let $F : \C \rightarrow \C$ be a strong functor and
assume $\pair\E\M$ is closed under $F$. If $\pair S\beta$ and $\pair
T{\beta'}$ are objects of $\Mnd{\C}{F}$ and $m$ is a
$\Mnd{\C}{F}$-morphism we equip $m[S]$ with a $\Mnd{\C}{F}$-object
structure $\pair {m[S]}{m[\beta]}$ by setting as below on the left. We
then have that $\epi m$ and $\mono m$ are $\Mnd{\C}{F}$-morphisms.
\begin{center}
  \begin{tikzcd}[row sep=5pt]
    |[alias=upperlt]|
    F\underline SX
    \arrow[r, "F\mono m_X" above, two heads]
    \arrow[d, "\beta_X" left]
    &
    F\underline{m[S]}X
    \arrow[d, "F\mono m_X" right]
    \arrow[ldd, "{m[\beta]}" above left, dashed]
    \arrow[ddl, "" name=diagonal, phantom]
    \\
    \underline SX
    \arrow[d, "\epi m_X" left, two heads]
    &
    F\underline TX
    \arrow[d, "\beta'_X" right]
    \\
    \underline{m[S]}X
    \arrow[r, "\mono m_X" below, tail]
    &
    |[alias=lowerrt]|
    \underline TX
    \arrow[phantom, "=" above, from=diagonal, to=upperlt]
    \arrow[phantom, "="      , from=diagonal, to=lowerrt]
  \end{tikzcd}
  \hfil
  \begin{tikzcd}[row sep=large]
    S \arrow[r, "m" above]
    \arrow[d, "f_1" left] &
    T
    \arrow[d, "f_2" right]
    \\
    S' \arrow[r, "m'" below] &
    T'
    \arrow[ul, "=", phantom]
  \end{tikzcd}
$\implies$
\begin{tikzcd}[row sep=large]
  S \arrow[r, two heads]
  \arrow[d, "f_1" left]
  \arrow[rd, "=", phantom] &
  m[S] \arrow[r, tail]
  \arrow[d, "{m[f]}" , dashed]
  &
  T
  \arrow[d, "f_2" right]
  \\
  S' \arrow[r, two heads] &
  m[S'] \arrow[r, tail]
  &
  T'
  \arrow[ul, "=" right, phantom]
\end{tikzcd}
\end{center}
Using the diagonal fill-in property, we can \emph{functorially}
factorise commuting squares of monad structure morphisms, i.e.,
morphisms $f = \pair {f_1}{f_2}$ between monad structure morphisms, as
above on the right.

\begin{theorem}[Functoriality]\label{theorem:monad-factorisation-functorial}
  Let $\pair{\E}{\M}$ be a factorisation system for a category $\C$,
  and let $f : \triple STm \to \triple {S'}{T'}{m'}$ be a commuting
  square of monad structure morphisms. If $\pair\E\M$ is closed under
  $S$ and $S'$, then $m[f] : m[S] \to m[S']$ is a monad structure
  morphism that preserves all of the above structure that $f$
  preserves:
  \begin{itemize}
  \item if $\pair\E\M$ is closed under products and $f$ is strong, then so is $m[f]$; and
  \item if moreover $f$ is an $F$-monad structure morphism, then
    $m[f]$ is an $F$-monad structure morphism.
  \end{itemize}
\end{theorem}

So far, we have worked with an arbitrary factorisation system
$\pair\E\M$. When it is an \emph{epi-mono} factorisation system, i.e.,
a pair $\pair\E\M$ in which $\M$ consists of monos, then
Theorem~\ref{theorem:monad-factorisation} holds under the weaker
assumption that $T$ is a monad, while $S$ need only be a monad
structure.  To prove it, instead of appealing to the diagonal fill-in
property, use the cancellation property of monos.

\subsection{Free monads}
To apply the Factorisation Theorem~\ref{theorem:monad-factorisation},
we need to choose a suitable monad $S$ and monad morphism $m$. When
giving semantics to type-and-effect systems, we take $S$ to be the
free monad for the functor $F_\epsilon$ from the end of
\S\ref{subsub:monads}. Here we give a sufficient condition for
$F_\epsilon$, or more generally, any functor $F$, to be compatible
with the factorisation system.

\begin{lemma}\label{lemma:monad compatibility}
  Let $\C$ be a category with $\kappa$-directed colimits, $\kappa$ a
  regular cardinal, $F : \C \to \C$ be a $\kappa$-ranked functor, and
  $\pair\E\M$ a factorisation system over $\C$. If $F$ is compatible
  with $\pair\E\M$, then the free $F$-monad $S_F$ is compatible with
  $\pair\E\M$.
\end{lemma}

To apply the last lemma to the signature functor $F_\epsilon$, we want
to show that $F_\epsilon$ preserves $\kappa$-directed colimits for
some $\kappa$, and that $\E$ is closed under $F_\epsilon$. For colimit
preservation, the following lemma covers our examples.
\begin{lemma}\label{lemma:signature functor compatibility}
  Let $\epsilon$ be an effect signature in a locally presentable
  cartesian closed category $\C$. Then the functor $F_\epsilon$
  preserves $\kappa$-directed colimits for some regular cardinal
  $\kappa$.
\end{lemma}
However, some $F_\epsilon$ may be incompatible with some factorisation
systems:
\begin{example}
Consider the (dense, full) factorisation system on
$\wCPO$. Exponentials ${(-)}^{Y}$ preserve dense maps iff $Y$ is a
countable discrete \wcpo. For a simple illustration, take the discrete
natural numbers $\naturals$ and the ordinal $\omega + 1$. Take $Y :=
\omega + 1$, and consider the inclusion $e : \naturals \to \omega +
1$, which is a dense map. Every monotone function $f : \omega + 1 \to
\naturals$ is constant, and so the \wchain{}-closure of
$e^Y[\naturals^Y]$ contains only constant functions. Therefore, the
identity function $x := \id \in (\omega + 1)^Y$ is not in this
closure, hence $e^Y$ isn't dense.
\end{example}

\section{Type-and-effect systems}\label{sec:calculus}
We consider a variant of
Moggi's~\cite{moggi:computational-lambda-calculus-and-monads}
computational $\lambda$-calculus, $\lambdac$, and its refinement with
a Gifford-style type-and-effect system. The denotational semantics for
such a system is standard, and we focus on the specific model
structure given by the Factorisation
Theorem~\ref{theorem:monad-factorisation}.

\subsection{Syntax}
The syntax of $\lambdac$ are parametrised by three sets: a set
$\BaseType$ of \emph{base types} ranged over by $b$; a set
$\underline\Sigma$ of \emph{operations} ranged over by $\op$; and a
set $\underline\Constant$ of \emph{constants} ranged over by $c$. We
also have the metavariable $x$ range over some set of variables and
$\epsilon$ ranges over finite subsets of $\underline\Sigma$. The
syntax of types $A, B$ (base types, products and sums, and function
types), ground types $G$, and terms $M$ of the $\lambdac$-calculus is
given as follows:
\begin{align*}
   &&
  M, N~ &::= c~|~\op\,M~|~x~|~\ttterm~|~(M, N)~|~\fstterm{M}~|~\sndterm{M}~|~\botelimterm{M}\\
  A, B ::=~& b~|~\unittype~|~A\hphantom{{}_1} \times B\hphantom{{}_2}~|~\emptytype~|~A\hphantom{{}_1} + B\hphantom{{}_2}~|~A \xrightarrow{\epsilon} B
  \!\!\!
  && \!\!\!\!|~\inlterm{M}~|~\inrterm{M}~|~\caseterm{M}{x}{N_1}{y}{N_2}~
  \\G ::=~& b~|~\unittype~|~G_1 \times G_2~|~\emptytype~|~G_1 + G_2 &
  &\!\!\!\!|~\lambda x.\,M~|~M\,N
\end{align*}
The main difference to Moggi's calculus is that we include a specified
set of constructs $\op\,M$ for causing effects. The other constructs
are standard: built-in constants, unit value, products with
projections, empty type elimination construct, sum injections and
pattern matching, and function abstraction and application.

To define $\lambdac$'s type system, we need some typing information
for effect operations and the constants.  Formally, a \emph{$\lambdac$
  signature} is a triple $\triple{\BaseType}{\Sigma}{\Constant}$
consisting of: a set $\BaseType$ of base types; a family of pairs of
ground type $\Sigma$ indexed by a set of operations
$\underline\Sigma$; and a family of types $\Constant$ indexed by a set
$\underline\Constant$.  We write $c : A$ when the type $A$ is the
$c$-component of $\Constant$, and $\op : G \rightarrow G'$ when $\pair
G{G'}$ is the $\op$-component of $\Sigma$.

\begin{figure}
  \[
  \begin{array}{@{}c@{}}
    \inferrule
      { (c : A) \in \Constant }
      { \effwelltyped{\Gamma}{c}{A}{\emptyset} }
    \quad
    \inferrule
      { \effwelltyped{\Gamma}{M}{A}{\epsilon} \\ (\op : A \to B) \in \Sigma }
      { \effwelltyped{\Gamma}{\op\,M}{B}{\epsilon\union\{\op\}} }
    \quad
    \inferrule
      { \effwelltyped{\Gamma}{M}{A}{\epsilon} \\       \epsilon \subset \epsilon'}
      { \effwelltyped{\Gamma}{{M}}{A}{\epsilon'}}
    \quad
    \inferrule
      { (x : A) \in \Gamma }
      { \effwelltyped{\Gamma}{x}{A}{\emptyset} }
    \quad
    \inferrule
      { }
      { \effwelltyped{\Gamma}{()}{A}{\emptyset} }
    \\[15pt]
    \inferrule
      { \effwelltyped{\Gamma}{M}{A}{\epsilon} \\
        \effwelltyped{\Gamma}{N}{B}{\epsilon'} }
      { \effwelltyped{\Gamma}{(M, N)}{A \times B}{\epsilon\union\epsilon'} }
    \quad
    \inferrule
      { \effwelltyped{\Gamma}{M}{A \times B}{\epsilon} }
      { \effwelltyped{\Gamma}{\fstterm{M}}{A}{\epsilon} }
    \quad
    \inferrule
      { \effwelltyped{\Gamma}{M}{A \times B}{\epsilon} }
      { \effwelltyped{\Gamma}{\sndterm{M}}{B}{\epsilon} }
    \quad
    \inferrule
      { \effwelltyped{\Gamma}{M}{0}{\epsilon} }
      { \effwelltyped{\Gamma}{\botelimterm{M}}{A}{\epsilon} }
    \\[15pt]
    \inferrule
      { \effwelltyped{\Gamma}{M}{A}{\epsilon} }
      { \effwelltyped{\Gamma}{\inlterm{M}}{A + B}{\epsilon} }
    \quad
    \inferrule
      { \effwelltyped{\Gamma}{M}{B}{\epsilon} }
      { \effwelltyped{\Gamma}{\inrterm{M}}{A + B}{\epsilon} }
    \quad
    \inferrule
      { \effwelltyped{\Gamma}{M}{A_1 + A_2}{\epsilon} \\
        \effwelltyped{\Gamma, x : A_1}{N_1}{B}{\epsilon'} \\
        \effwelltyped{\Gamma, y : A_2}{N_2}{B}{\epsilon'} }
      { \effwelltyped{\Gamma}{\caseterm{M}{x}{N_1}{y}{N_2}}{B}{\epsilon\union\epsilon'} }
    \\[15pt]
    \inferrule
      { \effwelltyped{\Gamma, x : A}{M}{B}{\epsilon} }
      { \effwelltyped{\Gamma}{\lambda x.\,M}{A \xrightarrow{\epsilon} B}{\emptyset} }
    \quad
    \inferrule
      { \effwelltyped{\Gamma}{M}{A \xrightarrow{\epsilon''} B}{\epsilon} \\
        \effwelltyped{\Gamma}{N}{A}{\epsilon'} }
      { \effwelltyped{\Gamma}{M\,N}{B}{\epsilon\union\epsilon'\union\epsilon''} }
  \end{array}
  \]
  \caption{$\lambdac$ type-and-effect system}
  \label{figure:lambdac-effect-typing}
\end{figure}

Given a $\lambdac$ signature we define two type systems. The
type-and-effect system consists of a typing judgment
$\effwelltyped{\Gamma}{M}{A}{\epsilon}$ given inductively by the rules
in Figure~\ref{figure:lambdac-effect-typing}.  Such judgements assert
that in \emph{typing context} $\Gamma$, a finitely supported partial function
from variable names to types, the term $M$ has type $A$ and uses only
the operations in $\epsilon \subset \underline\Sigma$. The rules are
standard for such systems.

We recover the usual type system for $\lambdac$ by erasing the
effect annotations $\epsilon$ from the type syntax and from
Figure~\ref{figure:lambdac-effect-typing}. In detail, for each type
$A$ there is an erased type $\erase{A}$, and similarly for contexts
$\Gamma$. The unrefined typing judgments
$\welltyped{\erase{\Gamma}}{M}{\erase{A}}$ are generated by the rules
of Figure~\ref{figure:lambdac-effect-typing} without annotations. This
judgment places no constraints on the operations that $M$ can use.  We
have that if $\effwelltyped{\Gamma}{M}{A}{\epsilon}$ then
$\welltyped{\erase{\Gamma}}{M}{\erase{A}}$.

\subsection{Semantics}\label{section:effect-system-semantics}
Fix a $\lambdac$ signature $\triple{\BaseType}{\Sigma}{\Constant}$. An
\emph{unrefined $\lambdac$ model}, consists of: a bicartesian closed
category $\C$; an object $\sem{b}\in\C$ for each $b\in\BaseType$; a
$\Sigma$-monad $T$ on $\C$; a Kleisli arrow $\sem\op : \sem G \to
T\sem{G'}$ for every $\op : G \to G'$ in $\Sigma$; and a morphism
$\sem{c} : 1 \rightarrow \sem{\erase{A}}$ for each constant $(c : A)
\in \Constant$. Unrefined models interpret the unrefined judgments
$\welltyped{\erase{\Gamma}}{M}{\erase{A}}$, with types and contexts denoting
$\C$-objects $\sem{\erase{B}}$ and $\sem{\erase{\Gamma}}$, and
judgements denoting Kleisli arrows
$\sem{\welltyped{\erase{\Gamma}}{M}{\erase{A}}} : \sem{\Gamma} \rightarrow
T\,\sem{\erase{A}}$.

To interpret type-and-effect judgements in their greatest generality,
one replaces the monad with a \emph{graded
  monad}~\cite{Katsumata:2014:PEM:2578855.2535846}. Here, as we
restrict to Gifford-style systems, we consider a simpler structure.  A
\emph{refined $\lambdac$ model} consists of: a bicartesian closed
category $\C$; an object $\sem{b}\in\C$ for each $b\in\BaseType$;
these data allow us to interpret effect annotations $\epsilon$ as
effect signatures $\sem\epsilon$, and so we further require a
functorial assignment $T_{-}$, to each $\epsilon\subseteq\Sigma$, of
an $\sem\epsilon$-monad $T_\epsilon$ on $\C$, and to each inclusion
$\epsilon\subseteq\epsilon'$ an $\sem\epsilon$-monad morphism
$T_\epsilon\to T_{\epsilon'}$; and a morphism $\sem{c} : 1 \rightarrow
\sem{A}$ for each constant $(c : A) \in \Constant$.  We interpret the
refined judgement $\effwelltyped{\Gamma}{M}{A}{\epsilon}$ by a
morphism $\sem{\Gamma} \to T_\epsilon\,\sem{A}$ along the same lines
of the unrefined semantics.

The main difference between the two model structures is the functorial
assignment $T_{-}$, which requires additional structure over the
unrefined model structure that is exponential in the number of
operations. We can derive it in the following way and under the
following assumptions, in addition to the unrefined model
structure. First, we assume that, for each $\epsilon \subset \Sigma$,
we have the free $\epsilon$-monad $S_{\epsilon}$. Second, we assume a
factorisation system $\pair\E\M$ that is closed under products and
each $S_{\epsilon}$. By Lemmata~\ref{lemma:monad compatibility}
and~\ref{lemma:signature functor compatibility} these two assumptions
hold in any locally presentable cartesian closed category in which
$\E$ is closed under exponentiation by the interpretation of base
types. Third, we assume a $\sem\Sigma$-monad $T$. For every $\epsilon
\subset \Sigma$, by initiality of the free $\epsilon$-monad, we have a
unique monad morphism $m_{\epsilon} : S_{\epsilon} \to T$. Applying
the Factorisation Theorem~\ref{theorem:monad-factorisation} to this
monad morphism, we set $T_{\epsilon} := m_{\epsilon}[S_{\epsilon}]$.
Applying the functorial action of $m_{\epsilon}[-]$ to the (unique)
$\epsilon$-monad morphism $S_{\epsilon \subset \epsilon'} :
S_{\epsilon} \to S_{\epsilon'}$, we set $T_{\epsilon \subset
  \epsilon'} := m_{\epsilon}[S_{\epsilon \subset
    \epsilon'}]$. Finally, we assume a refined interpretation of the
built-in constants compatible with this structure.

\subsection{Example reasoning}\label{subsec:reasoning}
We demonstrate the model construction on a small set-theoretic example.  Let $\Loc$
be a finite set of global memory location names. For our $\lambdac$
signature, we take:
\(
\BaseType := \set{\calc{Loc}, \calc{int}}
\), \(
\Sigma := \set{\getop : \calc{Loc} \to \calc{int}, \setop : \calc{Loc}\times\calc{int} \to 1}
\), and \[
\Constant := \set{+ : \smash{\calc{int}\times\calc{int} \xto{\emptyset} \calc{int}}}
\union \set{\calc{loc} : \calc{Loc} \suchthat \calc{loc} \in \Loc}
\union \set{a : \calc{int} \suchthat a \in \integers}
\]
For the unrefined model structure, we interpret: \(
\sem{\calc{Loc}} := \Loc
\) and \(
\sem{\calc{int}} := \integers
\).
For our monad, we set $\State := \integers^{\Loc}$ and take $T$ to be
the $\State$-state monad, $TX := (\State \times X)^{\State}$, with the
usual interpretation for $\getop$ and $\setop$. We interpret locations
and integers as themselves, and $+$ as addition without side effects.

For the refined model, we take the (surjection, injection)
factorisation system on $\Set$. We can calculate that
$T_{\set{\setop}}X = (\terminal + \integers)^{\Loc}\times X$ is the
writer monad for the following \emph{overwriting monoid}
$\triple{(\terminal + \integers)^{\Loc}}{\mathbf 1}{*}$:
\[
\mathbf 1 := \seq[\ell \in \Loc]{\injection_1 \star}
\qquad
\parent{\seq[\ell \in \Loc]{a_{\ell}} * \seq[\ell \in \Loc]{b_\ell}}_{\ell'} = \begin{cases}
  b_{\ell'} & b_{\ell'} \neq \injection_1\star \\
  a_{\ell'} & \text{otherwise}
\end{cases}
\]
I.e., an injected unit value at location $\ell$ represents no state
change, while an injected integer $a$ represents an update of that
location to $a$.  To see why, first note that the free
$\set{\setop}$-monad is the smallest set satisfying $S_{\set{\setop}}X
\isomorphic X + \Loc\times \integers \times S_{\set{\setop}}X$. The
unique $\set{\setop}$-monad morphism $m_{\set{\setop}} :
S_{\set{\setop}} \to T$ satisfies:
\[
m_{\set\setop}(\injection_1x) := \lambda s.\pair sx
\qquad
m_{\set\setop}(\injection_2\triple \ell ar) := \lambda s.\pair {s[\ell \mapsto a]}{m_{\set\setop}(r)}
\]
Factorising it, and using the finiteness of $\Loc$, we get the surjection:
\[
\epi m_{\set\setop}(\injection_1x) \mapsto \pair{\injection_1\star}{x}
\quad
\epi m_{\set\setop}(\injection_2\triple \ell ar) \mapsto {((\seq[\ell' \in \Loc]{\injection_1\star}[\ell \mapsto \injection_2a]*(-))\times \id)}\parent{m_{\set\setop}(r)}
\]
We then interpret $+$ as addition, as $T_{\emptyset}$ is the identity
monad. We can then validate the example from the introduction,
i.e. in the refined semantics $\sem{M+M} = \sem{(\lambda x.x+x) M}$ for every
$\effwelltyped{\Gamma}{M}{\calc{int}}{\set{\setop}}$.

\section{Monadic lifting}\label{sec:lifting}
To prove that the refined factorisation semantics matches the
unrefined semantics we use a suitable notion of logical relation. In
this section we define a notion of \emph{factorisation system for
  logical relations}, and show that these systems induce a suitable
logical relation. This notion combines Hughes and
Jacobs's~\cite{hughes-jacobs:factorization-systems-fibrations}
characterisation of fibrations arising from factorisation systems with
Katsumata's~\cite{katsumata:relating-computational-effects} fibrations
for logical relations.

\subsection{Preliminaries}

First we review some standard properties of fibrations, see
Jacobs~\cite{jacobs:categorical-logic} for a systematic development of
fibred category theory in type theory and logic. Instead of
considering general fibrations, we will only consider the simpler case
of faithful fibrations.

Let $p : \D \to \C$ be a faithful functor. For all $\D$-objects $X$,
$Y$, we write $f : X \rightarrowdot Y$ when $f : p\,X \to p\,Y$ in
$\C$ and there is some (necessarily unique) $\dot{f} : X \to Y$ such
that $p\,\dot{f} = f$. In this case we say that $f$ \emph{lifts} to
$\dot{f}$. If $f : X \rightarrowdot Y$ then $\dot{f}$ is
\emph{Cartesian} when, for all objects $Z\in\D$ and $g : p\,Z \to
p\,X$ with $f\compose g : Z \rightarrowdot X$ we have $g :
Z\rightarrowdot X$. The functor $p$ is a \emph{fibration} when, for
every object $Y$ in $\D$ and morphism $f : I \to p\,Y$ in $\C$ there
is an object $X$ such that $p\,X = I$ and $f : X\rightarrowdot Y$ is
Cartesian.

If $p : \D \to \C$ is a faithful fibration, we view objects $X\in\D$
as predicates over $X$, and morphisms $\dot{f} : X\rightarrow Y$ as
truth-preserving maps. If $f : p\,X \to p\,Y$ then $f :
X\rightarrowdot Y$ means $f$ is truth-preserving, and $\dot{f}$ is a
witness to this preservation. Faithfulness implies that $\dot{f}$ is
unique, so constructing such witnesses amounts to checking a property,
instead of providing structure. Cartesianness of $\dot{f}$ intuitively
means that $X$ is true on as many elements of $p\,X$ as possible, with
the constraint that $f$ is truth-preserving.

For every $I\in\C$, the \emph{fibre} $\Fibre{\D}{I}$ is the category
consisting of objects $X\in\D$ such that $p\,X = I$ and morphisms $f :
X \to Y$ in $\D$ such that $p\,f = \id{}_I$. We write $X\leq Y$ when
there is a (necessarily unique) morphism from $X$ to $Y$ in
$\Fibre{\D}{I}$, and $X \equiv Y$ when $X\leq Y$ and $Y\leq X$.

For each $f : I \to J$ in $\C$ there is an inverse image functor
$\inverse{f} : \Fibre{\D}{J}\to\Fibre{\D}{I}$ that sends an object $X$
to an object $Y$ such that $f : X \rightarrowdot Y$ is Cartesian.
Such $Y$ is unique up to $\equiv$: for any $Y'$ with the same property
we have $Y \equiv Y'$. We will also postulate that $\inverse{f}$ has a
left adjoint $\direct{f} : \Fibre{\D}{I}\to\Fibre{\D}{J}$, the
\emph{direct image functor}. When $\direct{f}$ exists, we call $p$ a
\emph{bifibration}.

For fibrations to give us logical relations, we also require both
categories to be bicartesian closed, and require $p$ to preserve the
bi-cartesian closed structure. For example, products in $\D$ allow us
to form logical relations over a product, and preservation of products
implies that this relation has the usual property of logical
relations. We will also want to form \emph{conjunctions/intersections}
of logical relations; these are given by products in fibres.

Katsumata combines all of these requirements into a single notion. A
\emph{fibration for logical
  relations}~\cite{katsumata:relating-computational-effects} over a
bicartesian closed category $\C$ is a faithful fibration $p : \D \to
\C$ such that:
\begin{itemize}
\item{$p$ is a bifibration: each inverse image functor
  $\inverse{f}$ has a left adjoint $\direct{f}$;}
\item{$\D$ is bicartesian closed, and $p$ strictly preserves the
  bicartesian closed structure; and}
\item{each fibre $\Fibre{\D}{I}$ has all small products, denoted
  $\bigwedge$.}
\end{itemize}
Our only deviation from Katsumata's definition is to require fibres to
be pre-orders instead of partial orders, due to our use of
non-\emph{strict} factorisation systems.

Recall also the change-of-base construction which allows us to
construct new fibrations for logical relations from existing
ones:
\begin{lemma}[Katsumata~{\cite[Proposition 6]{katsumata:relating-computational-effects}}]\label{lemma:logrel-pullback}
  Let $p : \D \to \C$ be a fibration for logical relations, and let $F
  : \C' \to \C$ be a product-preserving functor.  The projection from
  the pullback $\inverse{F}\,p$ of $p$ along $F$ is a fibration for
  logical relations on $\C'$.
  \[\begin{tikzcd}
    \inverse{F}\,\D
    \arrow[r]
    \arrow[rd, "\lrcorner" pos=0.1, phantom]
    \arrow[d, "\inverse{F}\,p" left] &
    \D
    \arrow[d, "p"] \\
    \C'
    \arrow[r, "F" below] &
    \C
  \end{tikzcd}\]
\end{lemma}
When we choose $F := (\times) : \C\times\C \to \C$, we call
$\inverse{F}\,\D$ the category of binary logical $p$-relations over
$\C$.

\subsection{Fibrations from factorisation systems}
Let $\pair{\E}{\M}$ be a factorisation system on $\C$. The codomain
functor $\cod : \M \to \C$ sends an $\M$-morphism $m : X
\rightarrowtail Y$ to its codomain $Y$, recalling that we view $\M$ as
a full subcategory of the arrow category $\C^{\rightarrow}$, so that objects are
$\M$-monos and morphisms are commutative squares. Cartesian morphisms
for $\cod$ are exactly pullback squares.  Given an $\M$-morphism $m :
X'\rightarrowtail Y'$ and a morphism $f : Y \to Y'$, we construct the
Cartesian morphism required in the definition by taking the pullback
of $m$ along $f$:
\[\begin{tikzcd}
  X \arrow[r]
  \arrow[d, "\inverse{f}\,m" left, tail]
  \arrow[rd, "\lrcorner" pos=0.1, phantom] &
  X' \arrow[d, "m", tail] \\
  Y \arrow[r, "f" below] &
  Y'
\end{tikzcd}\]
$\inverse{f}\,m$ is necessarily in $\M$ due to the diagonal fill-in property.
Hence if $\C$ has all pullbacks then $\cod$ is a fibration. If this is
the case then $\cod$ is also a bifibration: the left adjoint
$\direct{f}$ takes an $\M$-morphism $m$ to the $\M$-morphism in the
factorisation of $f \compose m$.

\begin{example}
Consider the (surjection,injection) factorisation for $\Set$. Every
injection $m : X \rightarrowtail Y$ is equal to the composition of an
inclusion $i$ and an isomorphism. In this case, we have $m \equiv
i$. This fact rephrases that an injection is, up to isomorphism in the
fibre, a subset $X \subseteq Y$. The direct image functor $\direct{f}$
of a function $f : Y \to Y'$ maps this subset to $\{f\,x~|~x\in
X\}\subseteq Y'$. The inverse image functor $\inverse{f}$ maps a
subset $X'\subseteq Y'$ to $\{x~|~f\,x\in X'\}\subseteq Y$.
\end{example}

\begin{example}
  Similarly for the (dense, full) factorisation for $\wCPO$, the
  full functions are the chain-closed subsets. Inverse images are the
  usual inverse images, but direct images are now the \wchain-closure
  of the direct image.
\end{example}

We extend the work of Hughes and
Jacobs~\cite{hughes-jacobs:factorization-systems-fibrations}, who give
a correspondence between certain factorisation systems (on categories
with pullbacks) and fibrations with additional properties. We restrict
this correspondence to fibrations for logical relations.

\begin{definition}[cf.~\cite{hughes-jacobs:factorization-systems-fibrations}]
  Let $\C$ be bicartesian closed. A factorisation system $\pair{\E}{\M}$ over $\C$ is a \emph{
    factorisation system for logical relations} when:
  \begin{multicols}{2}
    \it
  \begin{itemize}
    \item{$\C$ has all pullbacks of $\M$-morphisms;}
    \item{every morphism in $\M$ is a monomorphism\\
      (i.e.\ $m \compose f = m \compose g \Rightarrow f = g$);}
    \item{for every $Y\in\C$ the fibre $\Fibre{\M}{Y}$ has small
      products;}
    \item{$\M$ is closed under binary coproducts; and}
    \item{$\E$ is closed under binary products.}
  \end{itemize}
  \end{multicols}
\end{definition}

The monomorphism requirement implies that $\cod$ is faithful. The closure
of $\M$ under coproducts implies that $\M$ is bicartesian (it
automatically has initial and terminal objects and products). The closure of
$\E$ under binary products implies that for $m' : X' \rightarrowtail Y'$
the canonical morphism
$\expdot{X}{m'} : \exp{X}{X'} \rightarrowtail \exp{X}{Y'}$ is an
$\M$-morphism, and hence that $\M$ has exponentials, which are given by the
following pullback:
\[\begin{tikzcd}
  Z
  \arrow[d, "\expdot{m}{m'}" left, tail]
  \arrow[rd, "\lrcorner" pos=0.1, phantom]
  \arrow[r] &
  \exp{X}{X'} \arrow[d, "\expdot{X}{m'}", tail] \\
  \exp{Y}{Y'} \arrow[r, "\expdot{m}{Y'}" below] & \exp{X}{Y'}
\end{tikzcd}\]

\begin{lemma}
  \label{lemma:logrel-characterization}
  Let $\pair{\E}{\M}$ be a factorisation system over a bicartesian
  closed category $\C$. The codomain functor $\cod : \M \rightarrow
  \C$ is a fibration for logical relations iff $\pair{\E}{\M}$ is a
  factorisation system for logical relations.
\end{lemma}
This lemma also has a converse: if a fibration for logical relations
is a \emph{factorisation
fibration}~\cite[Definition 3.1]{hughes-jacobs:factorization-systems-fibrations}
then the induced factorisation system is a factorisation system for
logical relations.

\begin{example}
  The factorisation systems (surjection, injection) for $\Set$ and
  (dense, full) for $\wCPO$ are factorisation systems for logical
  relations. If $\pair{\E}{\M}$ is a factorisation system for logical
  relations on $\C$, then (component-wise $\E$, component-wise $\M$)
  is a factorisation system for logical relations on
  $\functors{\W}{\C}$.
\end{example}

\subsection{Folklore lifting for algebraic operations}
Since our semantics uses monads, we also need to lift monads to the
category of logical relations.  Let $p : \D \to \C$ be a faithful
fibration, $\epsilon$ is an effect signature in $\D$, and $T$ be a
$p\,\epsilon$-monad on $\C$, where $p\,\epsilon$ is the effect
signature with operations $\op : p\,X \to p\,Y$ for $(\op : X \to Y)
\in \epsilon$.
A \emph{lifting} of $T$ to $\D$ is an $\epsilon$-monad on $\D$ such
that:
\begin{multicols}{2}
  \begin{itemize}
    \item{for each $X \in \D$ we have $p\,(\dot{T}\,X) = T\,(p\,X)$;}
    \item{for each $f : X \rightarrow Y$ we have
      $p\,(\dot{T}\,f) = T\,(p\,f)$;}
    \item{the unit lifts: $p\,(\return^{\dot T}) = \return^T$;}
    \item{the multiplication lifts: $p\,(\mult^{\dot T}) = \mult^T$;}
    \item{the strength lifts: $p\,(\strength^{\dot T}) = \strength^T$; and}
    \item{each $\op \in \epsilon$ lifts:
      $p\,(\dot{\alpha}_{\op}) = \alpha_{\op}$.\vphantom{$\return^{\dot T}$}}
  \end{itemize}
\end{multicols}
Only the object action of $\dot T$ is a required structure, the other
requirements are properties we need to check.

As each logical relations proof involving monads involves a lifting,
these occur in abundance, and usually in an ad-hoc fashion.  Two
general lifting techniques are
$\top\top$-lifting~\cite{katsumata:tt-lifting} and the codensity
lifting~\cite{katsumata:codensity-lifting}. The particular lifting we
use is the \emph{free lifting}, which is the $\epsilon$-monad that is
initial amongst all $\epsilon$-liftings. The construction of this
lifting is folklore, and is described for binary relations over $\Set$
in Kammar's thesis~\cite{DBLP:phd/ethos/Kammar14}. We describe it for
the general case of a fibration for logical relations here.

Let $p : \D \to \C$ be a fibration for logical relations with
\emph{essentially small fibres}, i.e. each fibre has a representing set of
objects up to $\equiv$. For each object $X \in \D$ define $\mathcal{R} X$
as the set of all $X'$ in the representing set of $\Fibre{\D}{T\,(p\,X)}$ such that:
\begin{center}
\begin{tabular}{@{}l@{ }l@{\qquad}l@{ }l@{}}
  \textbullet& The unit respects $X'$: $\eta : X \rightarrowdot X'$.  &
  \textbullet& For each $(\op : A \to B)\in\epsilon$ the algebraic
  operation $\alpha_\op$ respects $X'$: \\&&& $\alpha_{\op} :
  \expdot{B}{X'} \rightarrowdot \expdot{A}{X'}$, where $\expdot{}{}$
  denotes exponentials in $\D$.
\end{tabular}
\end{center}
This definition makes essential use of the bijection between algebraic
operations and Kleisli arrows, as the former localises the closure
condition to $X'$ alone.
The elements of $\mathcal{R} X$ can be thought of as candidates for
$\dot{T}\,X$. We define the free lifting of $T$ to $\D$ on objects by:
\(\dot{T}\,X \definedby \bigwedge \mathcal{R} X\), i.e.,
$\dot{T}\,X$ is the least element of
$\mathcal{R}\,X$ with respect to the order $\leq$ in the fibre. This
definition extends uniquely to an $\epsilon$-monad $\dot{T}$ on $\D$.
\begin{theorem}
  \label{theorem:monad-lifting}
  $\dot{T}$ is a lifting of $T$ to $\D$, and is initial: for all
  liftings $\dot{T'}$, the identity lifts to a (necessarily unique)
  $\epsilon$-monad morphism $\dot{T}\rightarrow\dot{T'}$.
\end{theorem}

\subsection{Completeness}\label{subsec:completeness}
We now return to the language $\lambdac$ and relate the refined
semantics we construct with the unrefined semantics.  Suppose that the
factorisation system we used to construct the refined semantics is a
factorisation system for logical relations that is
\emph{well-powered}: each object has a representing set of
$\M$-morphisms into it, and let $p : \LogRel \to \C\times\C$ be the
fibration for logical relations constructed from the codomain
fibration $\cod : \M \to \C$, as in Lemma~\ref{lemma:logrel-pullback}.
Explicitly, an object of $\LogRel$ is a triple $\triple{X}{Y}{m}$
where $m : Z \rightarrowtail X \times Y$ (for some $Z$) is an
$\M$-mono. The \emph{diagonal} relations are the objects
$\triple{X}{X}{\delta_X}$, where $\delta_X = \langle{\id}, {\id}\rangle : X
\rightarrowtail X\times X$. We further assume that all diagonal
relations exist, i.e., the diagonals $\delta_X$ are in
$\M$. Well-poweredness of the factorisation system implies $\cod$ has
essentially  small fibres.
% Also let
% $R : \LogRel \to \M$ be the functor that takes a relation to its
% underlying $\M$-mono (so that $R\,\triple{X}{Y}{m} = m$.

\begin{example}
  The factorisation systems (surjection, injection) over $\Set$ and
  (dense, full) over $\wCPO$ are well-powered and have all
  diagonals. For every factorisation system $\pair\E\M$ for $\C$ and
  every small category $\W$, the factorisation (component-wise $\E$,
  component-wise $\M$) is well-powered if $\pair\E\M$ is well-powered,
  and has diagonals if $\pair\E\M$ has diagonals.
\end{example}

\begin{example}
  Over $\Set$, the factorisation system (iso, any) is not
  well-powered, and the factorisation system (any, iso) does not have
  all diagonals.
\end{example}

Consider any unrefined model together with a refined factorisation
model for it. For each $\epsilon\subseteq\Sigma$ both $T$ and
$T_\epsilon$ are $\epsilon$-monads, so $\pair{T_\epsilon}{T}$ is an
$\epsilon$-monad on $\C\times\C$ (and this forms a refined $\lambdac$
model on $\C\times\C$). By Theorem~\ref{theorem:monad-lifting} we can
lift $\pair{T_\epsilon}{T}$ to get an $\epsilon$-monad
$\dot{T}_\epsilon$ on $\LogRel$. Moreover, each monad morphism
$T_{\epsilon\subseteq\epsilon'}$ induces an $\epsilon$-monad morphism
$\dot{T}_\epsilon \to \dot{T}_\epsilon$. If we take the
interpretations $\LogRel{\sem{b}}$ of base types $b$ to be diagonal
relations $\triple{\sem{b}}{\sem{b}}{\delta_{\sem{b}}}$, we need to
interpret the constants to form a refined $\lambdac$ model on
$\LogRel$. By the fibration's faithfulness, this interpretation is
merely a property, and not a structure we need to provide. Using an
inductive argument, ground types $G$ denote diagonal
relations, and if $p\,(\LogRel\sem{c})$ is the interpretation of the
constant $c$ in $\C\times\C$ then for all well-typed terms
$\effwelltyped{\Gamma}{M}{A}{\epsilon}$ we have:
\[p\,(\LogRel\sem{\effwelltyped{\Gamma}{M}{A}{\epsilon}})
= \pair{\sem{\effwelltyped{\Gamma}{M}{A}{\epsilon}}}
{\sem{\welltyped{\erase{\Gamma}}{M}{\erase{A}}}}
\]
We use $\LogRel$ to compare the refined model we constructed with the
original unrefined model. First:

\begin{lemma}
  \label{lemma:logrel-morphism-mono}
  Suppose that the initial $\epsilon$-monad $S_\epsilon$ is given by
  the transfinite construction from \S\ref{subsub:monads}. For each
  morphism $\pair{f_1}{f_2} : \pair{1}{1} \to
  \pair{T_\epsilon\,X}{T\,X}$ in $\C\times\C$, if $\pair{f_1}{f_2} : 1
  \rightarrowdot \dot{T}_\epsilon\triple{X}{X}{\delta_X}$ then $f_2 =
  \mono m_{\epsilon} \compose f_1$.
\end{lemma}

Under the combined assumptions of this subsection, we can now show
that the refined semantics is complete for equational reasoning. For
all closed terms of ground type $\effwelltyped{}{M}{G}{\epsilon}$ and
$\effwelltyped{}{N}{G}{\epsilon}$,
  \[
    \sem{\welltyped{}{M}{G}} = \sem{\welltyped{}{N}{G}}
    \quad\iff\quad
    \sem{\effwelltyped{}{M}{G}{\epsilon}} =
    \sem{\effwelltyped{}{N}{G}{\epsilon}}
  \]

To prove it, noting that ground types are interpreted as diagonal
relations, we apply Lemma~\ref{lemma:logrel-morphism-mono} to both
$\LogRel\sem{M}$ and $\LogRel\sem{N}$ to show that
\begin{gather*}
  \sem{\welltyped{}{M}{G}} = n \compose \sem{\effwelltyped{}{M}{G}{\epsilon}}
  \qquad
  \sem{\welltyped{}{N}{G}} = n \compose \sem{\effwelltyped{}{N}{G}{\epsilon}}
\end{gather*}
Now the result follows from the fact that every $\M$-morphism is a
monomorphism.

\section{Examples}\label{sec:examples}
Before we conclude, we apply the factorisation construction to several
examples.

\begin{example}
  Continuing the global state example from \S\ref{subsec:reasoning},
  we have the full factorisation:
\begin{center}
  $T_\emptyset = \Id$
  \hfil
  $T_{\{\getop\}} = \exp{\State}{(-)}$
  \hfil
  $T_{\{\setop\}} = (1 + \integers)^{\Loc}\times (-)$
  \hfil
  $T_{\{\getop, \setop\}} = T$
\end{center}
By the completeness of the refined semantics from
\S\ref{subsec:completeness}, we can apply the equation from
\S\ref{subsec:reasoning} to closed programs of ground type without
changing their denotations in the unrefined semantics.
\end{example}

\begin{example}
If instead of $T$ we use the monad
$\exp{\exp{(\exp{\exp{\State}{(-)}}{R})}{\State}}{R}$, which combines
global state with continuations (so that the language can include
constants such as call/cc) then we get the same factorisation,
assuming $|R| > 1$.  Hence we can also verify the caching
transformation in this situation.  The construction in Kammar's
thesis~\cite{DBLP:phd/ethos/Kammar14} does not allow this
factorisation, as it is restricted to Lawevere theories, i.e., ranked
monads, and $T$ is not ranked. Note that, as call/cc is not algebraic,
we cannot interpret call/cc in the refined semantics, so cannot
validate transformations on subprograms that use continuations.
\end{example}

\begin{example}
  Using the (dense, full) factorisation of $\wCPO$, we can re-cast
  Kammar and Plotkin's~\cite{Kammar:2012:AFE:2103656.2103698}
  validation of effect-dependent optimisations.
\end{example}

\begin{example}
Let $\valuetype$ be a base type for values of a ground type (with
associated constants). Consider $\lambdac$ with a base type $\reftype$
of references, and the set $\Sigma = \{\lookup : \reftype \to
\valuetype, \update : \reftype\times\valuetype \to \unittype, \alloc :
\valuetype \to \reftype\}$ of operations, so that we can read from and
write to references, and allocate new references. Let $\I$ be the
category of finite ordinals and injections between them. Plotkin and
Power~\cite{plotkin-power:notions-of-computation} interpret these
operations the functor category $\functors{\I}{\Set}$ as follows. Let
$\V$ be a nonempty finite set $\V$ of values with interpretation for
the $\valuetype$ constants. Then we interpret $\valuetype$ as the
constantly-$\V$ functor, and $\reftype$ as the Yoneda embedding
$\sem{\reftype} = \I(1, -)$, so that $\sem{\reftype}\,n$ has $n$
elements. The local state monad is defined using a coend:
\[T\,X\,n \definedby \exp{\V^n}{\coend{m \in \I}{\I(n, m) \times \V^m \times X m}}\]
A computation is given an initial state in $\V^n$, and returns
an injection that describes how the original $n$ references are
distributed over the $m$ references (so that $n \leq m$), a new state in
$\V^m$, and a result in $X m$.

The category $\functors{\I}{\Set}$ has a (pointwise surjection,
pointwise injection) factorisation system. For each subset
$\epsilon\in\Sigma$, since $\V$ is finite, we can show that the transfinite sequence
$S_\alpha$ converges at $\aleph_0$. We can therefore show by induction
on $\alpha$ that, for example, there are component-wise surjections from the
corresponding free monads into the following functors:
\begin{align*}
  T_{\{\alloc\}}\,X\,n \definedby \coend{m \in \I}{\I(n, m) \times \V^{m - n} \times X m} & &
  T_{\{\lookup, \update\}}\,X\,n \definedby \exp{\V^n}{X n \times \V^n}
\end{align*}
Calculation shows that there are pointwise injections from these into
$T$. Theorem~\ref{theorem:monad-factorisation} (and the uniqueness of
factorisations) implies they are the monads that result from
factorisation. Now note that there are two sequencing morphisms
$T_{\{\alloc\}}\,X \times T_{\{\alloc\}}\,Y \to T_{\{\alloc\}}\,(X
\times Y)$, one that does the left computation first and one that does
right first.  It is easy to check that these are equal, i.e.,
$T_{\set\alloc}$ is commutative, and hence we can validate a
transformation that reorders computations that only allocate.
\end{example}

%% \OK{Sum examples: include Paul's theorem for Set and example for wCpo}
%% \OK{Tensor examples: including Eckmann-Hilton and non-determinism (from Ohad's thesis)}

\section{Conclusion}\label{sec:conclusion}
We have presented a factorisation theorem for cutting down a monad
into sub-monads based on a factorisation system. We showed how this
construction gives uniform semantics for Gifford-style type-and-effect
systems. Synthesising Hughes and Jacobs's characterisation of
fibrations arising from factorisation systems and Katsumata's
axiomatisation of fibrations for logical relations, we provide a
general proof that the factorisation construction is sound and
complete for effect-dependent equational reasoning.

We would like to generalise the completeness theorem to programs of
higher-order types, and not just ground
types. Reynolds~\cite{reynolds:continuations} relates direct and
continuation semantics by defining domain-theoretic partial maps
between the two semantics, and proves such a theorem. Felleisen and
Cartwright~\cite{Cartwright1994} provide an analogous construction and
proof for free effects and their
handlers~\cite{plotkin-pretnar:effect-handlers-journal,BAUER2015108},
but their semantics does not involve monads. Well-powered factorisation systems for
logical relations induce categories of partial maps via Fiore's
axiomatic domain theory~\cite{fiore:book}. The axiomatic development
is particularly appealing because factorisation systems of interest,
such as the (dense, full) factorisation of $\wCPO$ do not admit a
representation using a lifting monad.

We want to relate the free lifting to other lifting techniques, most
notably $\top\top$-, and codensity-, lifting. We would also like to
relate Benton et al.'s~\cite{BENTON201827} relational models to our
construction. We want to apply this construction to more sophisticated
computational effects, such as dynamic memory
allocation~\cite{8005109}. Another application area to the free
lifting is relational parametricity with effects --- we have used it
as a semantic precursor to the more syntactic work on analysing the
value restriction~\cite{kammar_pretnar_2017}, and we hope it applies
more widely. Finally, there is still a wide gap between Gifford-style
type-and-effect systems and the full generality of graded monads. We
hope our account would carry over to such settings.

\inpar{Acknowledgements}
This work has been supported by an Engineering and Physical Sciences
Research Council (ESPRC) studentship, EPSRC grants EP/N007387/1
`Quantum computation as a programming language', EPSRC Leadership
Fellowship EP/H005633/1 `Semantic Foundations for Real-World Systems',
Institute for Information \& Communications Technology Promotion
(IITP) grant funded by the Korea government (MSIP) No.~R0190-16-2011
`Development of Vulnerability Discovery Technologies for IoT Software
Security', a Balliol College Oxford Career Development Fellowship,
European Research Council Grant `Events, Causality and Symmetry ---
the next generation semantics', and Isaac Newton Trust grant `algebraic
theories, computational effects, and concurrency'. We would like to
thank Marcelo Fiore, Mathieu Huot, Justus Matthiesen, and Ian Orton
for fruitful discussions and suggestions.

\bibliographystyle{plain}
\bibliography{felleisen-wadler}

\begin{thebibliography}{10}

\bibitem{BAUER2015108}
Andrej Bauer and Matija Pretnar.
\newblock Programming with algebraic effects and handlers.
\newblock {\em Journal of Logical and Algebraic Methods in Programming},
  84(1):108 -- 123, 2015.
\newblock Special Issue: The 23rd Nordic Workshop on Programming Theory (NWPT
  2011) Special Issue: Domains X, International workshop on Domain Theory and
  applications, Swansea, 5-7 September, 2011.

\bibitem{BENTON201827}
Nick Benton, Martin Hofmann, and Vivek Nigam.
\newblock Effect-dependent transformations for concurrent programs.
\newblock {\em Science of Computer Programming}, 155:27 -- 51, 2018.
\newblock Selected and Extended papers from the International Symposium on
  Principles and Practice of Declarative Programming 2016.

\bibitem{benton-kennedy-russell:compiling-standard-ml-to-java-bytecodes}
Nick Benton, Andrew Kennedy, and George Russell.
\newblock Compiling {Standard ML} to {Java Bytecodes}.
\newblock In {\em Proceedings of the Third ACM SIGPLAN International Conference
  on Functional Programming}, ICFP '98, pages 129--140, New York, NY, USA,
  1998. ACM.

\bibitem{bousfield:constructions-of-factorization-systems-in-categories}
{Aldridge K.} Bousfield.
\newblock Constructions of factorization systems in categories.
\newblock {\em Journal of Pure and Applied Algebra}, 9(2):207--220, 1977.

\bibitem{Cartwright1994}
Robert Cartwright and Matthias Felleisen.
\newblock {\em Extensible denotational language specifications}, pages
  244--272.
\newblock Springer Berlin Heidelberg, Berlin, Heidelberg, 1994.

\bibitem{fiore:book}
Marcelo~Pablo Fiore.
\newblock {\em Axiomatic Domain Theory in Categories of Partial Maps}.
\newblock Distinguished Dissertations in Computer Science. Cambridge University
  Press, 1996.

\bibitem{hughes-jacobs:factorization-systems-fibrations}
Jesse Hughes and Bart Jacobs.
\newblock Factorization systems and fibrations.
\newblock {\em Electronic Notes in Theoretical Computer Science}, 69:156 --
  182, 2003.

\bibitem{jacobs:categorical-logic}
Bart Jacobs.
\newblock {\em Categorical Logic and Type Theory}.
\newblock Number 141 in Studies in Logic and the Foundations of Mathematics.
  North Holland, Amsterdam, 1999.

\bibitem{8005109}
O.~Kammar, P.~B. Levy, S.~K. Moss, and S.~Staton.
\newblock A monad for full ground reference cells.
\newblock In {\em 2017 32nd Annual ACM/IEEE Symposium on Logic in Computer
  Science (LICS)}, pages 1--12, June 2017.

\bibitem{DBLP:phd/ethos/Kammar14}
Ohad Kammar.
\newblock {\em Algebraic theory of type-and-effect systems}.
\newblock PhD thesis, University of Edinburgh, {UK}, 2014.

\bibitem{Kammar:2012:AFE:2103656.2103698}
Ohad Kammar and Gordon~D. Plotkin.
\newblock Algebraic foundations for effect-dependent optimisations.
\newblock In {\em Proceedings of the 39th Annual ACM SIGPLAN-SIGACT Symposium
  on Principles of Programming Languages}, POPL '12, pages 349--360, New York,
  NY, USA, 2012. ACM.

\bibitem{kammar_pretnar_2017}
Ohad Kammar and Matija Pretnar.
\newblock No value restriction is needed for algebraic effects and handlers.
\newblock {\em Journal of Functional Programming}, 27:e7, 2017.

\bibitem{katsumata:tt-lifting}
Shin-ya Katsumata.
\newblock A semantic formulation of {$\top$}{$\top$}-lifting and logical
  predicates for computational metalanguage.
\newblock In Luke Ong, editor, {\em Computer Science Logic}, pages 87--102,
  Berlin, Heidelberg, 2005. Springer Berlin Heidelberg.

\bibitem{katsumata:relating-computational-effects}
Shin-ya Katsumata.
\newblock Relating computational effects by {$\top$}{$\top$}-lifting.
\newblock {\em Inf. Comput.}, 222:228--246, 2013.

\bibitem{Katsumata:2014:PEM:2578855.2535846}
Shin-ya Katsumata.
\newblock Parametric effect monads and semantics of effect systems.
\newblock {\em SIGPLAN Not.}, 49(1):633--645, 2014.

\bibitem{katsumata:codensity-lifting}
Shin-ya Katsumata and Tetsuya Sato.
\newblock Codensity liftings of monads.
\newblock In {\em CALCO}, 2015.

\bibitem{kelly:transfinite-constructions}
G.M. Kelly.
\newblock A unified treatment of transfinite constructions for free algebras,
  free monoids, colimits, associated sheaves, and so on.
\newblock {\em Bulletin of the Australian Mathematical Society}, 22(1):1--83,
  1980.

\bibitem{kelly:transfinite-constructions-addenda}
G.M. Kelly.
\newblock Two addenda to the author's ‘transfinite constructions’.
\newblock {\em Bulletin of the Australian Mathematical Society},
  26(2):221–237, 1982.

\bibitem{lehmann-pasztor:epis-need-not-be-dense}
Daniel Lehmann and Ana Pasztor.
\newblock Epis need not be dense.
\newblock {\em Theoretical Computer Science}, 17(2):151 -- 161, 1982.

\bibitem{lucassen-gifford:polymorphic-effect-systems}
John~M. Lucassen and David~K. Gifford.
\newblock Polymorphic effect systems.
\newblock In {\em Proceedings of the 15th ACM SIGPLAN-SIGACT Symposium on
  Principles of Programming Languages}, POPL '88, pages 47--57, New York, NY,
  USA, 1988. ACM.

\bibitem{maclane:working-mathematician}
Saunders Mac~Lane.
\newblock {\em Categories for the Working Mathematician (Graduate Texts in
  Mathematics)}.
\newblock Springer, 2nd edition, 1998.

\bibitem{marmolejo-wood:monads-as-extension-systems}
Francisco Marmolejo and Richard~J. Wood.
\newblock Monads as extension systems --- no iteration is necessary.
\newblock {\em Theory and Applications of Categories}, 24(4):84--113, 2010.

\bibitem{meseguer:factorisations}
Jos\'e Meseguer.
\newblock Completions, factorizations and colimits for $\omega$-posets.
\newblock In {\em Mathematical Logic in Computer Science, Salgotarjan, 1978,
  Colloquia Mathematica Societatis Janos Bolyai}, volume~26, pages 509--545.
  North Holland, 1981.

\bibitem{moggi:computational-lambda-calculus-and-monads}
Eugenio Moggi.
\newblock Computational lambda-calculus and monads.
\newblock In {\em Proceedings of the Fourth Annual Symposium on {L}ogic in
  {C}omputer {S}cience}, pages 14--23, Piscataway, NJ, USA, 1989. IEEE Press.

\bibitem{plotkin-power:algebraic-operations-generic-effects}
Gordon Plotkin and John Power.
\newblock Algebraic operations and generic effects.
\newblock {\em Applied Categorical Structures}, 11(1):69--94, 2003.

\bibitem{plotkin-power:notions-of-computation}
Gordon~D. Plotkin and John Power.
\newblock Notions of computation determine monads.
\newblock In {\em Proceedings of the 5th International Conference on
  Foundations of Software Science and Computation Structures}, pages 342--356,
  London, UK, 2002. Springer-Verlag.

\bibitem{plotkin-pretnar:effect-handlers-journal}
Gordon~D. Plotkin and Matija Pretnar.
\newblock Handlers of algebraic effects.
\newblock In {\em Proceedings of the 18th European Symposium on Programming
  Languages and Systems: Held As Part of the Joint European Conferences on
  Theory and Practice of Software, ETAPS 2009}, ESOP '09, pages 80--94.
  Springer-Verlag, Berlin, Heidelberg, 2009.

\bibitem{power1999enriched}
A~John Power.
\newblock Enriched {L}awvere theories.
\newblock {\em Theory and Applications of Categories}, 6(7):83--93, 1999.

\bibitem{reynolds:continuations}
John~C. Reynolds.
\newblock On the relation between direct and continuation semantics.
\newblock In Jacques Loeckx, editor, {\em Automata, Languages and Programming},
  pages 141--156, Berlin, Heidelberg, 1974. Springer.

\bibitem{tolmach:hierarchy}
Andrew~P. Tolmach.
\newblock Optimizing {ML} using a hierarchy of monadic types.
\newblock In {\em Types in Compilation}, pages 97--115, 1998.

\bibitem{Wadler:1998:MEM:291251.289429}
Philip Wadler.
\newblock The marriage of effects and monads.
\newblock {\em SIGPLAN Not.}, 34(1):63--74, 1998.

\end{thebibliography}

\end{document}